\documentclass[11pt,twoside]{article}  
\usepackage{adassconf}

\begin{document}   

%
%

\paperID{O5-1}

%
%
%
%
\def\la{\mathrel{\hbox{\rlap{\hbox{\lower4pt\hbox{$\sim$}}}\hbox{$<$}}}}
\def\ga{\mathrel{\hbox{\rlap{\hbox{\lower4pt\hbox{$\sim$}}}\hbox{$>$}}}}
\title{Science with Virtual Observatory Tools}

%

\author{P.\ Padovani\altaffilmark{1}}
\affil{European Southern Observatory, Karl-Schwarzschild-Str. 2, 
       D-85748 Garching bei M\"unchen, Germany, 
       Email Paolo.Padovani@eso.org}


\altaffiltext{1}{Astrophysical Virtual Observatory Scientist}


\contact{Paolo Padovani}
\email{Paolo.Padovani@eso.org}

%
%

\paindex{Padovani, P.}

%
%

\authormark{Padovani}




\begin{abstract}          
The Virtual Observatory is now mature enough to produce cutting-edge
science results. The exploitation of astronomical data beyond classical
identification limits with interoperable tools for statistical
identification of sources has become a reality. I present the discovery of
68 optically faint, obscured (i.e., type 2) active galactic nuclei (AGN)
candidates in the two GOODS fields using the Astrophysical Virtual
Observatory (AVO) prototype. Thirty-one of these sources have high
estimated X-ray powers ($>10^{44}$ erg/s) and therefore qualify as
optically obscured quasars, the so-called QSO 2. The number of these
objects in the GOODS fields is now 40, an improvement of a factor $> 4$
when compared to the only 9 such sources previously known. By going $\sim
3$ magnitudes fainter than previously known type 2 AGN in the GOODS fields
the AVO is sampling a region of redshift -- power space much harder to
reach with classical methods. I also discuss the AVO move to our next
phase, the EURO-VO, and our short-term plans to continue doing science with
the Virtual Observatory.
\end{abstract}


\section{Astronomy in the XXI century}
\setcounter{footnote}{1}

Astronomy is facing the need for radical changes. When dealing with surveys
of up to $\sim 1,000$ sources, one could apply for telescope time and
obtain an optical spectrum for each one of them to identify the whole
sample. Nowadays, we have to deal with huge surveys (e.g., the Sloan
Digital Sky Survey [\htmladdnormallinkfoot{SDSS}{http://www.sdss.org/}],
the Two Micron All Sky Survey
[\htmladdnormallinkfoot{2MASS}{http://www.ipac.caltech.edu/2mass/}], the
Massive Compact Halo Object
[\htmladdnormallinkfoot{MACHO}{http://wwwmacho.anu.edu.au/}] survey),
reaching (and surpassing) the 100 million objects. Even at, say, 3,000
spectra at night, which is only feasible with the most efficient
multi-object spectrographs and for relatively bright sources, such surveys
would require more than 100 years to be completely identified, a time which
is clearly much longer than the life span of the average astronomer! But
even taking a spectrum might not be enough to classify an object. We are in
fact reaching fainter and fainter sources, routinely beyond the typical
identification limits of the largest telescopes available (approximately 25
magnitude for 2 - 4 hour exposures), which makes ``classical''
identification problematic. These very large surveys are also producing a
huge amount of data: it would take more than two months to download at 1
Mbytes/s (a very good rate for most astronomical institutions) the Data
Release 3 (\htmladdnormallinkfoot{DR3} {http://www.sdss.org/dr3/}) SDSS
images, about a month for the catalogues. The images would fill up $\sim$
1,300 DVDs ($\sim$ 650 if using dual-layer technology). And the final SDSS
will be about twice as large as the DR3. These data, once downloaded, need
also to be analysed, which requires tools which may not be available
locally and, given the complexity of astronomical data, are different for
different energy ranges. Moreover, the breathtaking capabilities and
ultra-high efficiency of new ground- and space-based observatories have led
to a ``data explosion'', with astronomers world-wide accumulating more than
one Terabyte of data per night (judging from some of the talks at this
conference, this is very likely to be an underestimate). For example, the
European Southern Observatory (ESO)/Space Telescope European Coordinating
Facility (ST-ECF) archive is predicted to increase its size by two orders
of magnitude in the next eight years or so, reaching $\approx 1,000$
Terabytes. Finally, one would like to be able to use all of these data,
including multi-million-object catalogues, by putting this huge amount of
information together in a coherent and relatively simple way, something
which is impossible at present.

All these hard, unescapable facts call for innovative solutions. For
example, the observing efficiency can be increased by a clever
pre-selection of the targets, which will require some ``data-mining'' to
characterise the sources' properties before hand, so that less time is
``wasted'' on sources which are not of the type under investigation. One
can expand this concept even further and provide a ``statistical''
identification of astronomical sources by using all the available,
multi-wavelength information without the need for a spectrum. The
data-download problem can be solved by doing the analysis where the data
reside. And finally, easy and clever access to all astronomical data
worldwide would certainly help in dealing with the data explosion and would
allow astronomers to take advantage of it in the best of ways.

\section{The Virtual Observatory}

The name of the solution is the Virtual Observatory (VO). The VO is an
innovative, evolving system, which will allow users to interrogate multiple
data centres in a seamless and transparent way, to make the best use of 
astronomical data. Within the VO, data analysis tools and models,
appropriate to deal also with large data volumes, will be made more
accessible. New science will be enabled, by moving Astronomy beyond
``classical'' identification with the characterisation of the properties of
very faint sources by using all the available information. All this will
require good communication, that is the adoption of common standards and
protocols between data providers, tool users and developers. This is being
defined now using new international standards for data access and mining
protocols under the auspices of the recently formed International Virtual
Observatory Alliance (\htmladdnormallinkfoot{IVOA}{http://ivoa.net}), 
a global collaboration of the world's astronomical communities.

One could think that the VO will only be useful to astronomers who deal
with colossal surveys, huge teams and Terabytes of data! That is not the
case, for the following reason. The World Wide Web is equivalent to having
all the documents of the world inside one's computer, as they are all
reachable with a click of a mouse. Similarly, the VO will be like having
all the astronomical data of the world inside one's desktop. That will
clearly benefit not only professional astronomers but also anybody
interested in having a closer look at astronomical data. Consider the
following example: imagine one wants to find {\it all} the observations of
a given source available in {\it all} astronomical archives in a given
wavelength range. One also needs to know which ones are in raw or processed
format, one wants to retrieve them and, if raw, one wants also to have
access to the tools to reduce them on-the-fly. At present, this is
extremely time consuming, if at all possible, and would require, even to
simply find out what is available, the use a variety of search interfaces,
all different from one another and located at different sites. The VO will
make all this possible very easily.

\section{The VO in Europe and the Astrophysical Virtual Observatory}

The status of the VO in Europe is very good. In addition to seven current
national VO projects, the European funded collaborative Astrophysical
Virtual Observatory initiative (\htmladdnormallinkfoot{AVO} {\tt
http://www.euro-vo.org}) is creating the foundations of a regional scale
infrastructure by conducting a research and demonstration programme on the
VO scientific requirements and necessary technologies. The AVO has been
jointly funded by the European Commission (under the Fifth Framework
Programme [FP5]) with six European organisations participating in a three
year Phase-A work programme. The partner organisations are ESO in Munich,
the European Space Agency, AstroGrid (funded by PPARC as part of the United
Kingdom's E-Science programme), the CNRS-supported Centre de Donnees
Astronomiques de Strasbourg (CDS) and TERAPIX astronomical data centre at
the Institut d'Astrophysique in Paris, the University Louis Pasteur in
Strasbourg, and the Jodrell Bank Observatory of the Victoria University of
Manchester. The AVO is the definition and study phase leading towards the
Euro-VO - the development and deployment of a fully fledged operational VO
for the European astronomical research community. A Science Working Group
was also established to provide scientific advice to the project.

\begin{figure}
\plotone{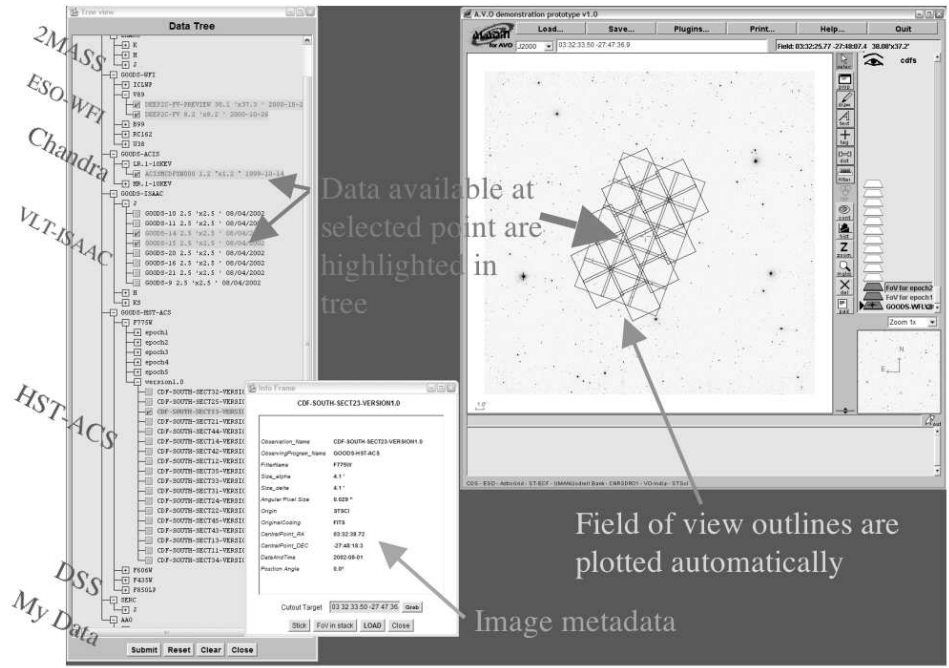}
\caption{The AVO prototype in action. An ESO/WFI image of the GOODS southern
field, overlaid with the HST/ACS data field of view outlines. The
``data-tree'' on the left shows the images available in the Aladin image
server. Data available at selected coordinates get highlighted in the tree.
Metadata information is also accessible. The user's own data can also be
loaded into the prototype. This is based on the use of IVOA agreed
standards, namely the Data Model, descriptive Metadata, and data
interchange standards.} \label{fig1}
\end{figure}

The AVO project is driven by its strategy of regular scientific
demonstrations of VO technology, held on an annual basis in coordination
with the IVOA. For this purpose progressively more complex AVO
demonstrators are being constructed. The current one, a downloadable Java
application, is an evolution of Aladin (\paperref{O5-2} \adassxiv),
developed at CDS, and has become a set of various software components,
provided by AVO and international partners, which allows relatively easy
access to remote data sets, manipulation of image and catalogue data, and
remote calculations in a fashion similar to remote computing (see
Fig. \ref{fig1}).

\section{Doing Science with the AVO}

The AVO held its second demonstration, 'AVO 1st Science', on January 27 -
28, 2004 at ESO. The demonstration was truly multi-wavelength, using
heterogeneous and complex data covering the whole electromagnetic
spectrum. These included: MERLIN, VLA (radio), ISO [spectra and images] and
2MASS (infrared), USNO, ESO 2.2m/WFI and VLT/FORS [spectra], and HST/ACS
(optical), XMM and Chandra (X-ray) data and catalogues. Two cases were
dealt with: an extragalactic case on obscured quasars, centred around the
Great Observatories Origin Deep Survey (GOODS) public data, and a Galactic
scenario on the classification of young stellar objects.

The extragalactic case was so successful that it turned into the first
published science result fully enabled via end-to-end use of VO tools and
systems, the discovery of $\sim 30$ high-power, supermassive black holes in
the centres of apparently normal looking galaxies. 

\section{Discovering optically faint, obscured quasars with VO tools}

How did we get a scientific paper out of a science demonstration? The
extragalactic science case revolved around the two GOODS fields (Giavalisco
et al. 2004a, \paperref{O3-1} \adassxiv), namely the Hubble Deep Field-North
(HDF-N) and the Chandra Deep Field-South (CDF-S), the most data-rich, deep
survey areas on the sky.  Our idea was to use the AVO prototype to look for
high-power, supermassive black holes in the centres of apparently normal
looking galaxies.

Black holes lurk at the centres of active galaxies (AGN) surrounded by dust
which is thought to be, on theoretical and observational grounds (see,
e.g., Urry \& Padovani 1995; Jaffe et al. 2004), distributed in a flattened
configuration, torus-like. When we look down the axis of the dust torus and
have a clear view of the black hole and its surroundings these objects are
called ``type 1'' AGN, and display the broad lines (emitted by clouds
moving very fast close to the black hole) and strong UV emission typical of
quasars. ``Type 2'' AGN, on the other hand, lie with the dust torus edge-on
as viewed from Earth so our view of the black hole is totally blocked by
the dust over a range of wavelengths from the near-infrared to soft X-rays.
The optical/UV spectrum of type 2 AGN is characterized by emission lines
much narrower than those of quasars, as they are emitted by clouds which
are further away and therefore move more slowly. 

While many dust-obscured low-power black holes, the Seyfert 2s, have been
identified, until recently few of their high-power counterparts were
known. This was due to a simple selection effect: when the source is a
low-power one and therefore, on average, closer to the observer, one can
very often detect some features related to narrow emission lines on top of
the emission from the host galaxy, which qualify it as a type 2 AGN. But
when the source is a high-power one, a so-called QSO 2, and therefore, on
average, further away from us, the source looks like a normal galaxy. Until
very recently, QSO 2s were selected against by quasar surveys, most of
which were tuned to find objects with very strong UV emission. The
situation has changed with the advent of Chandra and XMM-Newton, which are
providing a sensitive window into the hard X-ray emission of AGN.

\subsection{The Method} 

The two key physical properties that we use to identify type 2 AGN
candidates are that they be obscured, and that they have sufficiently high
power to be classed as an AGN and not a starburst. Our approach was to look
for sources where nuclear emission was coming out in the hard X-ray band,
with evidence of absorption in the soft band, a signature of an obscured
AGN, and the optical flux was very faint, a sign of absorption. One key
feature was the use of a correlation discovered by Fiore et al. (2003)
between the X-ray-to-optical ratio and the X-ray power, which allowed us to
select QSO 2s even when the objects were so faint that no spectrum, and
therefore no redshift, was available.

We selected absorbed sources by using the Alexander et al. (2003) X-ray
catalogues for the two GOODS fields, which provide counts in various X-ray
bands. We define the hardness ratio $HR = (H-S)/(H+S)$, where $H$ is the
hard X-ray counts ($2.0 - 8.0$ keV) and $S$ is the soft X-ray counts ($0.5
- 2.0$ keV). Szokoly et al. (2004) have shown that absorbed, type 2 AGN are
characterized by $HR \ge -0.2$. We adopt this criterion and identify those
sources which have $HR \ge -0.2$ as absorbed sources. We find 294 (CDF-S:
104, HDF-N: 190) such absorbed sources which represent $35^{+3}_{-2}\%$ of
the X-ray sources in the Alexander catalogues. Note that increasing
redshift makes the sources softer (e.g., at $z = 3$ the rest-frame $2 - 8$
keV band shifts to $0.5 - 2$ keV) so our selection criterion will
mistakenly discard some high-z type 2 sources, as pointed out by Szokoly et
al. (2004). The number of type 2 candidates we find has therefore to be
considered a lower limit.

The optical counterparts to the X-ray sources were selected by
cross-match\-ing the absorbed X-ray sources with the GOODS ACS catalogues
(29,599 sources in the CDF-S, 32,048 in the HDF-N). We used version v1.0
of the reduced, calibrated, stacked, and mosaiced images and catalogues as
made available by the \htmladdnormallinkfoot{GOODS
team}{http://www.stsci.edu/science/goods/}. The GOODS catalogues contain
sources that were detected in the $z$-band, with $BVi$ photometry in
matched apertures (Giavalisco et al. 2004b). 

We initially searched for optical sources that lay within a relatively
large threshold radius of 3.5\arcsec\ (corresponding to the maximal
$3\sigma$ positional uncertainty of the X-ray positions) around each X-ray
source. This was done using the cross match facility in the AVO prototype
tool using the ``best match'' mode. Since the 3.5\arcsec\ radius is large
relative to the median positional error, and given the optical source
density the initial cross match inevitably includes a number of false and
multiple matches. To limit our sample to good matches, we use the criterion
that the cross match distance be less than the combined optical and X-ray
$3\sigma$ positional uncertainty for each individual match. Applying this
distance/error $<1$ criterion we limit the number of matches to 168 (CDF-S:
65, HDF-N: 103). These matches are all within a much smaller radius than
our initial 3.5\arcsec\ threshold, with most of the distance/error $<1$
matches being within 1.25\arcsec~(and two matches at 1.4 and 1.5\arcsec).
The estimated number of false matches we expect to have is small, between 8
and 15\%.

Previously classified sources and their spectroscopic redshifts are
available from Szokoly et al. (2004) for the CDF-S and Barger et al. (2003)
for the HDF-N. Derivation of X-ray powers for these objects is
straightforward\footnote{Throughout this paper we adopt a cosmological
model with $H_0 = 70$ km s$^{-1}$ Mpc$^{-1}$, $\Omega_{\rm M} = 0.3$, and
$\Omega_{\rm \Lambda} = 0.7$}. For the unclassified sources we estimated
the X-ray power as follows: we first derived the $f(2 - 10 keV)/f(R)$ flux
ratio (converting the ACS $i$ magnitudes to the $R$ band), and then
estimated the X-ray power from the correlation found by Fiore et al.
(2003), namely $\log L_{2 - 10} = log f(2 - 10 keV)/f(R) + 43.05$ (Fiore,
p.c.; see their Fig. 5). Note that this correlation has an r.m.s. of $\sim
0.5$ dex in X-ray power. We stress that our estimated X-ray powers reach
$\sim 10^{45}$ erg/s and therefore fall within the range of the Fiore et
al. (2003) correlation. On the other hand it should be pointed out that our
sources are much fainter than the objects which have been used to calibrate
the Fiore et al.'s correlation.

The work of Szokoly et al. (2004) has shown that absorbed, type 2 AGN are
characterized by $HR \ge -0.2$. It is also well known that normal galaxies,
irrespective of their morphology, have X-ray powers that reach, at most,
$L_{\rm x} \la 10^{42}$ erg/s (e.g., Forman, Jones \& Tucker 1994; Cohen
2003). Therefore, any X-ray source with $HR \ge -0.2$ and $L_{\rm x} >
10^{42}$ erg/s should be an obscured AGN. Furthermore, following Szokoly et
al. (2004), any such source having $L_{\rm x} > 10^{44}$ erg/s will qualify
as a type 2 QSO.

\subsection{Results}

\begin{figure}
\epsscale{0.6}
\plotone{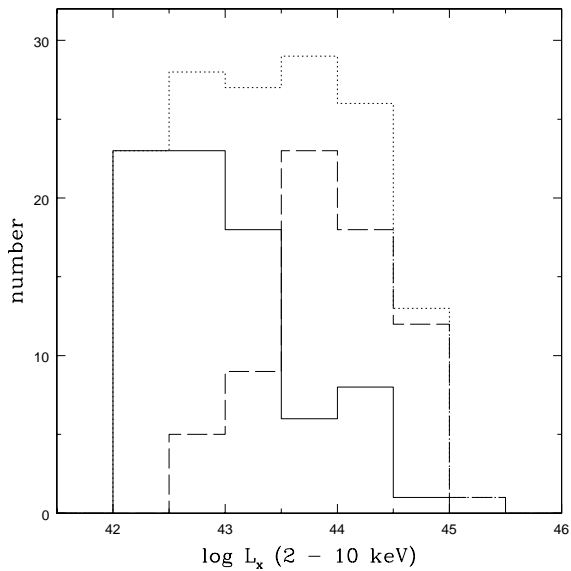}
\caption{The X-ray power distribution for our new type 2 AGN candidates
(dashed line), previously known type 2 AGN (solid line), and the sum of the
two populations (dotted line). QSO 2 are defined, somewhat arbitrarily, as
having $L_{\rm 2- 10 keV} > 10^{44}$ erg/s.}\label{fig2}
\end{figure}

Out of the 546 X-ray sources in the GOODS fields, 203 are absorbed ($HR \ge
-0.2$). Out of these we selected 68 type 2 AGN candidates, 31 of which
qualify as QSO 2 (estimated X-ray power $> 10^{44}$ erg/s). We note that
the distribution of estimated X-ray power covers the range $5 \times
10^{42} - 2 \times 10^{45}$ erg/s and peaks around $10^{44}$ erg/s (see
Fig. \ref{fig2}). The number of QSO 2 candidates, therefore, is very
sensitive to the dividing line between low- and high-luminosity AGN, which
is clearly arbitrary and cosmology dependent. For example, if one defines
as QSO 2 all sources with $L_{2 - 10} > 5 \times 10^{43}$ erg/s, a value
only a factor of 2 below the commonly used one and corresponding to the
break in the AGN X-ray luminosity function (Norman et al. 2002), the number
of such sources increases by $\sim 50\%$. We also note that, based on the
r.m.s.  around the Fiore et al. (2003) correlation, the number of QSO 2
candidates fluctuates in the $13 - 54$ region. The number of type 2 AGN, on
the other hand, can only increase, as all our candidates have estimated
$\log L_{2 - 10} > 42.5$.

Our work brings to 40 the number of QSO 2 in the GOODS fields, an
improvement of a factor $\sim 4$ when compared to the only nine such
sources previously known. As expected, being still unidentified, our
sources are very faint: their median ACS $i$ magnitude is $\sim 25.5$,
which corresponds to $R \sim 26$ (compare this to the $R \sim 22$ typical
of the CDF-S sources with redshift determination). The QSO 2 candidates are
even fainter, with median $i$ magnitude $\sim 26.3$ ($R \sim 26.8$).
Therefore, spectroscopical identification is not possible, for the large
majority of objects, even with the largest telescopes currently available.
We have used our estimated X-ray powers together with the observed fluxes
to derive redshifts for our type 2 candidates (tests we have performed on
the type 2 sources with spectroscopic redshifts show that this method,
although very simple, is relatively robust). Our type 2 AGN are expected to
be at $z \approx 3$, while our QSO 2 should be at $z \approx 4$. By using
VO methods we are sampling a region of redshift - power space so far much
harder to reach with classical methods. For the first time, we can also
assess how many QSO 2 there are down to relatively faint X-ray fluxes. We
find a surface density $> 330$ deg$^{-2}$ for $f(0.5 - 8 keV) \ge 10^{-15}$
erg cm$^{-2}$ s$^{-1}$, higher than previously estimated.

Fig. \ref{fig2} shows the X-ray power distribution for our new type 2 AGN
candidates (dashed line), previously known type 2 AGN (solid line), and the
combined sample (dotted line). It is interesting to note how the distributions
are very different, with the already known type 2 AGN peaking around $L_{\rm
x} \sim 10^{43}$ erg/s and declining for luminosities above $\sim 3 \times
10^{43}$ erg/s, while our new candidates are rising in this range and peak
around $L_{\rm x} \sim 10^{44}$ erg/s. To be more quantitative, while only
$\sim 1/5$ of already known type 2 AGN have $\log L_{\rm x} > 43.5$, $\sim
3/4$ of our candidates are above this value. This difference is easily
explained by our use of the X-ray-to-optical flux ratios to estimate X-ray
powers and by the fact that our candidates are on average $\sim 3$ magnitudes
fainter than previously known sources. Our method is then filling a gap in the
luminosity distribution, which becomes almost constant in the range $10^{42}
\la L_{\rm x} \la 3 \times 10^{44}$ erg/s. This also explains the fact that
the number of QSO 2 candidates we find is $\ga 3$ times larger than the
previously known ones. 

The identification of a population of high-power obscured black holes and
the active galaxies in which they live has been a key goal for astronomers
and will lead to greater understanding and a refinement of the cosmological
models describing our Universe. The paper reporting these results has been
recently published (Padovani et al. 2004).


The AVO prototype made it much easier to classify the
sources we were interested in and to identify the previously known ones, as
we could easily integrate all available information from images, spectra,
and catalogues at once. This is proof that VO tools have evolved beyond the
demonstration level to become respectable research tools, as the VO is
already enabling astronomers to reach into new areas of parameter space
with relatively little effort. 

The AVO prototype can be downloaded from the \htmladdnormallinkfoot{AVO Web
site}{http://www.\-euro-vo.\-org/twiki/bin/view/Avo/SwgDownload}. We
encourage astronomers to download the prototype, test it, and also use it
for their own research. For any problems with the installation and any
requests, questions, feedback, and comments you might have please contact
the AVO team at twiki@euro-vo.org. (Please note that this is still a
prototype: although some components are pretty robust some others are not.)

\section{Near Future AVO Science Developments}

The AVO is promoting science with VO tools through two further developments:
a Science Reference Mission and the next science demonstration. 

\subsection{The AVO Science Reference Mission}

The AVO team, with input from the Science Working Group, is putting
together a Science Reference Mission. This will define the key scientific
results that the full-fledged EURO-VO should achieve when fully implemented
and will consist of a number of science cases covering a broad range of
astronomical topics, with related requirements, against which the success
of the EURO-VO will be measured.

\subsection{The next AVO Science Demonstration}

The next and last AVO science demonstration is to be held in January 2005
at the European Space Astronomy Centre (ESAC; formerly known as VILSPA).
Preparations are still on-going so the details are not fully worked out yet
but it is firmly established that we will be dealing with two scenarios.
The first, on star formation histories in galaxies, will revolve around the
European Large-Area ISO Survey (ELAIS), which covers five different areas
of the sky over $\sim$ 10 deg$^2$. The second, on the transition from
Asymptotic Giant Branch to Planetary Nebulae, will be the strongest one on
the science side and should produce a new list of stars in this very
interesting transitional phase.

On the technical side, the science demonstration will see the rollout of
the first version of the EURO-VO portal, through which European astronomers
will gain secure access to a wide range of data access and manipulation
capabilities. Also, we will demonstrate the use of distributed workflows,
registry harvesting, and the wrapping of sophisticated astronomical
applications as Web services. 

The AVO demonstration will also mark the transition from the AVO to the
EURO-VO. Funding for the technology part of the EURO-VO, VO-TECH, has been
secured from the European Community at a level of 6.6 million Euros, which
will translate into 12 Full Time Equivalent (FTEs). Twelve more FTEs will
be provided by the partners, which include Edinburgh, Leicester, and
Cambridge in the UK, ESO, CDS, and INAF in Italy. 

\section{Summary}

The main results of this paper can be summarized as follows:

\begin{itemize}

\item The Virtual Observatory is happening because it has to! If it does
not, we will not be able to cope with the huge amount of data astronomers
are being flooded with.

\item Astronomy can and {\it is being} done with Virtual Observatory tools,
which are now mature enough. Real science results are being produced and
papers are being published. 

\item The Astrophysical Virtual Observatory, soon to be EURO-VO, is committed
to the pursuit of science with Virtual Observatory tools through scientific
demonstrations, science papers, and a Science Reference Mission. 

\end{itemize}

\acknowledgments

The obscured quasar paper was done in collaboration with Mark Allen, Piero
Rosati, and Nic Walton. It is a pleasure to thank the AVO team for their
superb work, without which the paper would have not been possible, and the
many people who have produced the data on which the paper is based,
particularly the GOODS, CDF-S, Penn State, and HELLAS Teams.

\end{document}